\documentstyle[twoside,fleqn,npb,epsfig]{article}
\newcommand{\AmS}{{\protect\the\textfont2
  A\kern-.1667em\lower.5ex\hbox{M}\kern-.125emS}}

\def    \be             {\begin{equation}}
\def    \ee             {\end{equation}}
\def    \beq             {\begin{equation}}
\def    \eeq             {\end{equation}}
\def    \ba             {\begin{eqnarray}}
\def    \ea             {\end{eqnarray}}
\def    \beqn           {\begin{eqnarray}}
\def    \eeqn           {\end{eqnarray}}
\def    \beeq           {\begin{eqnarray}}
\def    \eeeq           {\end{eqnarray}}

\def    \=              {\;=\;}
\def    \frac           #1#2{{#1 \over #2}}

\def    \bra#1          {\mbox{$\langle #1 |$}}
\def    \ket#1          {\mbox{$| #1 \rangle$}}
\def    \to             {\rightarrow} 

\newcommand     \MSB            {\ifmmode {\overline{\rm MS}} \else 
                                 $\overline{\rm MS}$  \fi}

\def    \as             {\ifmmode \alpha_s \else $\alpha_s$ \fi}

\def\rt1{\raisebox{-1ex}{\rlap{$\; \rho \to 1 \;\;$}}
\raisebox{.4ex}{$\;\; \;\;\simeq \;\;\;\;$}}

\def\b0{\beta_0}

\def\ban{\begin{eqnarray*}}
\def\ean{\end{eqnarray*}}
\def\nno{\nonumber \\}

  \newcommand{\ccaption}[2]{
    \begin{center}
    \parbox{0.40\textwidth}{
      \caption[#1]{\small{{#2}}}
      }
    \end{center}
    }

\def    \zeit   #1#2#3{{Z. Phys.} {\bf C#1}  (19#2) #3}
\def    \np     #1#2#3{{Nucl. Phys.} {\bf B#1} (19#2) #3}
\def    \journal #1#2#3#4{{\it #1} {\bf #2} (19#3) #4}   
\def    \pr     #1#2#3{{Phys. Rev.} {\bf D#1}  (19#2) #3}
\def    \prl    #1#2#3{{Phys. Rev. Lett.} {\bf #1}  (19#2) #3}

\title
{ \vskip -20pt
\hfill {\small INFNFE-03-00} \\
\vskip 15pt
A new approach to multi-jet 
calculations in hadron collisions}

\author{ 
M.~Moretti\address{Department of Physics, University of Ferrara and
INFN sezione di Ferrara} }

\begin{document}

\begin{abstract} 

The 
ALPHA
algorithm to evaluate the exact, tree-level matrix elements 
is 
reviewed in
the context of  multi-parton processes in QCD. 
The algorithm is suited
for the authomatic calculation of tree-level scattering amplitudes
and allows for straightforward inclusion of mass effects.
It's CPU cost grows like $K^n$ ($K$ constant, $n$ number of external particles)
as opposed to the factorial growth of Feynman graphs.
 I also discuss
the  summation over colour configurations 
which is designed to allow
the construction of parton-level event generators suitable to interfacing with
a parton-shower evolution including the effects of colour-coherence.
Explicit results for the total rates and differential distributions of
processes with 8 final-state partons are given.
\end{abstract}

\maketitle

\section{Introduction}

The ability to
evaluate production rates for multi-jet final states will be
fundamental at the LHC to study a large class of processes, within and
beyond the SM. A
 necessary
feature of any multi-jet calculation is the possibility to properly
evolve the purely partonic final state, for which exact fixed-order
perturbative calculations can be performed, into the observable
hadronic final state. This evolution is best performed using shower
Monte Carlo calculations. The accurate description of color-coherence
effects, furthermore, requires 
a
careful bookkeeping of the contribution to the matrix elements of all
possible color configurations. The goal of the 
algorithm~\cite{Caravaglios:1999yr}
described in this talk is to allow the effective calculation of
multi-parton matrix elements, allowing the separation, to the leading
order in $1/N_c^2$ ($N_c=3$ being the number of colors), of the
independent color configurations. This technique allows an
unweighting of the color configurations, and allows the merging of
the parton level calculation with the {\tt HERWIG} Monte Carlo.

The key element of the strategy is the use of the algorithm {\tt
ALPHA}, introduced in Ref.~\cite{Caravaglios:1995cd} for the evaluation of
arbitrary multi-parton matrix elements. The algorithm is 
built up from an {\em iteration} of matrix multiplication and is
intended
for the {\em authomatic} calculation of tree-level amplitudes.
It has a complexity growing like a
power in the number of particles, compared to the factorial-like
growth that one expects from naive diagram counting. 

The {\tt
ALPHA} algorithm will be reviewed in
section (\ref{alphasec}), and its application to multi-jets physics in
section (\ref{nj}).

\section {The {\tt
ALPHA} algorithm }
\label{alphasec}

In reference \cite{Caravaglios:1995cd}, a new approach to the computation 
of tree level scattering amplitudes was introduced.
This approach, based on the numerical  Legendre transform of the effective 
action, is  particularly useful for the 
authomatic calculation of multi-particle 
processes. 
This technique was implemented in a {\tt FORTRAN} code 
\cite{Caravaglios:1995cd} which has                                       
been succesfully used  to study several intricated electroweak processes 
\cite{alpha-appllep,alpha-applnlc} of interest
both at {\tt LEP} \cite{alpha-appllep} and at the {\tt NLC} 
\cite{alpha-applnlc}. 
Among the key features of the algorithm are:
\begin{itemize}

\item Suitable (easy to implement) 
for the authomatic calculation of the scattering amplitudes.

\item Inclusion of mass effects is straightforward 

\item The CPU cost  grows like $K^n$ ($K$ constant, 
$n$ number of external particles)
as opposed to the factorial growth of the
number of Feynman graphs.

\end{itemize} 
We review in this Section 
the {\tt ALPHA} algorithm; the interested reader can find                  
a more detailed discussion in the 
original  paper \cite{Caravaglios:1999yr,Caravaglios:1995cd},
 which includes
an explicit analytic example for the $\lambda \phi^3 $ theory.

Let $\Gamma$ be the one-particle-irreducible generator of the 
Green functions for a given theory. Then 
the computation of the S-matrix requires the evaluation 
of  the  Legendre transform, Z, of $\Gamma$:
\[                                                      
Z(J^\alpha)= - \Gamma(\phi^\alpha) 
+ J^\alpha(x)\phi^\alpha(x)     
\label {legtr}
\]
where $\phi^\alpha$ are the classical fields defined as the solutions of
\[                                                                  
 J^\alpha =\frac  {\delta \Gamma }{\delta \phi^\alpha} \; ,
\label{minim}                                              
\]
and the $J^\alpha$  
play the role of classical sources.

For concretness we will develop in some detail
the case of the scattering amplitue
for $n$ external gluons.
At tree-level $\Gamma$ coincide with the Lagrangian.
In momentum space
\ban
{\cal L}_{YM}(A) & = & -1/2 (p_\mu A^a_\nu-p_\nu A^a_\mu)^2
\\
& & + g f_{abc}(p_\mu A^a_\nu-p_\nu A^a_\mu)A_\mu^bA_\nu^c\\
& &  -
(B^a_{\mu\nu})^2 - 2 g f_{abc}B^a_{\mu\nu}A_\mu^b A_\nu^c + J_\mu^a A_\mu^a
\ean
where $B^a_{\mu\nu}$ is an auxiliary field and it is introduced
in order to deal with trilinear interactions only.
The sources
$J_\mu^a$ is a  standard source term: $J(p)=\sum_{j=1}^n
\epsilon^a_\mu\delta(p-p_j)$, {\em i.e.}~it 
contains the relevant excitation for
the external particles. Notice that because of this choice only
a finite number of momenta (linear combinations of $p_j$)
enters into the problem which is reduced to a problem
with a finite number of degrees of freedom.

The fields $A(p)$ are then
found as solutions  of the 
equation of motion. Let us stick to Feynman gauge for definitness.
\ba
A_\mu^a(p)
 & = & \frac{g}{p^2}  f_{abc}\left [ 2( k-p) \cdot A^b(q) A_\mu^c(k) 
\right .
\nno
& &
- 
2 q_\mu  A^b(q)\cdot A_\mu^c(k) 
\nno
& & 
\left . - B^b_{\mu\nu}(q) A_\nu^c(k)
\right ] + \frac{1}{p^2} J_\mu(p)
\nno
B^a_{\mu\nu}(p) 
& = & - g f_{abc} A_\mu^b(q) A_\nu^c(k)
\nno
\label{motion}
\ea
where momentum conservation is understood.

 The  equations of motion (\ref{motion})
 are solved {\em iteratively} (expansion in $g$)
and this implies that 
the problem is solved with a {\em loop of matrices multiplication},
more suitable for numerical implementation than the standard approach.
The initialization steps are
\ba
 A_\mu^a (p_j)
& = & 
\epsilon_\mu^a (p_j)  
\nno
 B^a_{\mu\nu}(p_j)
& = & 
0
\nno
\label{init}
\ea 
and the subsequent steps
\ba
A_\mu^a (p_j+p_k) & = & 
\nno
& & \hskip -50pt  \frac{g f_{abc}}{(p_j+p_k)^2} \left [ - 2 (p_j+ 2 p_k) 
\cdot A^b(p_j) A_\mu^c(p_k) 
\right .  
\nno
& & 
\hskip 5pt
+  2 (p_k+ 2 p_j) 
\cdot A^b(p_k) A_\mu^c(p_j) 
\nno
 & & \hskip 5pt
- 2  (p_j+p_k)_\mu  A^b(p_k) \cdot A^c(p_j) 
\nno
 & & \hskip 5pt
\left . 
- B^b_{\mu\nu}(p_j) A_\mu^c(p_k)
\right ]
\nno
 B^a_{\mu\nu}(p_j+p_k)
& = &
-g f_{abc} A^b(p_k) A_\mu^c(p_j)
\nno
\label{iter}
\ea
The step (\ref{iter}) is then iterated until
we have constructed $A$ and $B$ fields with
up to $n$ momenta.
An important remark is in order here. Performing the
iteration (\ref{iter}) we drop terms containing $\epsilon^a(p_j)
 \epsilon^a(p_j)$ {\em i.e.~}twice the same external momenta. In 
fact, although legitimate, they don't contribute to
the amplitude.

Finally the amplitude
is \be
{\cal A}( p_{j_1},\dots , p_{j_n}) = {\cal L}_{YM}(A)
\label {ampl}
\ee
where the $A$ fields are obtained in eqns.~(\ref{init},\ref{iter})\footnote
{Using the equation of motion it is possible to halve
the required number of iteration step as well as to reduce
the number of contributions entering the final
expression of the amplitude.}.
Notice that the prescription to drop terms containing at least
twice the same external momenta ($\sim \epsilon^a(p_j)
 \epsilon^a(p_j)$) is still kept and it is because of this prescription
toghether with the choice (\ref{init}) for the initial step that
neither truncation nor functional derivation is actually required.

A remark is in order here. The algorithm sketched
above has an important feature: it provides a very compact 
way of storing the relevant information. Indeed
the number of contraction $A^a(p_{j_1}+\dots+p_{j_m})$
(see eq.~\ref{iter}) which is needed is of the order
$2^n$, each of them requiring roughly the same CPU time to
be computed. Therefore the CPU cost of the algorithm
grows like a {\em constant to the power n,
n} being the number of external particles,
as opposed to the factorial growth of
the number of Feynman graphs.
To hint how does this work let us develop in more detail
the case of the amplitude for five external gluons.
For simplicity we neglect the four gluons coupling.
The $A$ fields are given by
\ba
 A_\mu^a (p_j)
& = & 
\epsilon_\mu^a (p_j)  \hskip 30pt j=1,\dots, 5
\nno
A_\mu^a (p_j+p_k) & = & 
\nno
& & \hskip -50pt \frac{g f_{abc}}{(p_j+p_k)^2} \left [ -2 (p_j+ 2 p_k) 
\cdot A^b(p_j) A_\mu^c(p_k) 
\right .  
\nno
& & 
\hskip 5pt
+ 2 (p_k+ 2 p_j) 
\cdot A^b(p_k) A_\mu^c(p_j) 
\nno
 & & 
\hskip 5pt \left . - 2  (p_j+p_k)_\mu  A^b(p_k) \cdot A^c(p_j) 
\right ]
\nno
& & \hskip 60pt j=1,\dots, 5
\nno
\label{suba}
\ea
and the {\tt ALPHA} amplitude by
\ba
{\cal A} & = &
g \sum_{j=1}^5  A^a(p_j)_\alpha 
\nno
& & 
\hskip -50pt \sum_{k\ne l\ne m\ne n\ne j}^5 f_{abc}
\tau_{\alpha\beta\gamma}
A^b(p_k+p_l)_\beta A^c(p_m+p_n)_\gamma 
\nno
\tau_{\alpha\beta\gamma}
& = &
g_{\alpha\beta}(p_j-p_k-p_l)_\gamma 
\nno
& &
+ g_{\gamma\alpha}(p_m+p_n-p_j)_\beta \nno
& & 
+g_{\beta\gamma}(p_k+p_l-p_m-p_n)_\alpha
\nno
\label {ampl5}
\ea
notice that 
the summations
over $k,l,m,n$,
as well as over lorentz and color indices,
 are carried over {\em before} the multiplication
by $ A^a(p_j)$. It is this feature extensively used
in the construction of both the $A$ fields and  the
amplitude which turns the factorial growth of the number of
Feynman graphs to a power law growth of the CPU cost
of the {\tt ALPHA} algorithm.
Notice that in eqns.~(\ref{suba},\ref{ampl5}) 
$A(p_j+p_k+p_m)$ and $A(p_j+p_k+p_m+p_l)$ 
are not computed (and used). 
Morover neither the kinetic term nor
the source one appears in the expression
(\ref{ampl5}) of the amplitude.
These simplifications
occur because of the equation of motion as
discussed in more detail in 
\cite{Caravaglios:1999yr,Caravaglios:1995cd}. 

\section {Multi-jets processes}
\label{nj}

Multi-jet final states play an important role in the study of high-energy
collisions. They provide in fact interesting signatures 
for several phenomena,
both within the Standard Model (e.g. top-pair production), and beyond it
(e.g. multi-jet decays of supersymmetric particles such as gluinos and
squarks).  The accurate determination of the properties of these phenomena
requires a good understanding of the properties of the usually large
multi-jet QCD backgrounds, 
which can distort the shapes of signal distributions
and affect the measurement of quantities such as resonances' masses.

There are several reasons 
for wanting to improve the tools currently available
to perform these calculations.           
\begin{enumerate}                             
\item First of all, interesting final states with larger jet multiplicities
will become available with the next generation of colliders (LHC and       
NLC).
\item Secondly, one would like to be able to complement the calculation of
parton-level matrix elements 
with the evaluation of the full hadronic structure
of the final state.
\end {enumerate} 

 The goal of the algorithm~\cite{Caravaglios:1999yr}
described in this Section is to allow the effective calculation of
multi-parton matrix elements, allowing the separation, to the leading
order in $1/N_c^2$ ($N_c=3$ being the number of colors), of the
independent color configurations. This technique allows an
unweighting of the color configurations, and allows the merging of
the parton level calculation with the {\tt HERWIG} Monte Carlo.
The key ingredient of our strategy is the {\tt ALPHA} algorithm
outlined in the previous section. It's power-law growing in complexity
is a necessary feature of any attempt to evaluate matrix elements for
processes with large numbers of 
external particles, since the number of Feynman
diagrams grows very quickly  (see table 1)
beyond any reasonable value.
\begin{table*}[hbt]
\setlength{\tabcolsep}{1.5pc}
\newlength{\digitwidth} \settowidth{\digitwidth}{\rm 0}
\catcode`?=\active \def?{\kern\digitwidth}
\begin{tabular*}{\textwidth}{@{}@
{\extracolsep{\fill}}|c|r|r|r|r|}
\hline                                                     
 \rule[-7 pt]{0 pt}{24 pt} 
Process                 & $n=7$ & $n=8$ & $n=9$ & $n=10$ \\
\hline                                                     
\hline
\rule[-7 pt]{0 pt}{24 pt}
$ g~g      \to n\; g$   & 559,405 & 
10,525,900 & 224,449,225 & 5,348,843,500 \\
\hline                                                               
\rule[-7 pt]{0 pt}{24 pt}
$ q \bar q \to n\; g$   & 231,280 & 
4,016,775 & 79,603,720 &  1,773,172,275 \\
\hline                                                             
\end{tabular*}
\caption{Number of Feynman diagrams corresponding to amplitudes with
different numbers of quarks and gluons.}
\end{table*}

Once the hard scattering matrix element is known the
 prescription \cite{Caravaglios:1999yr}
 to correctly generate the parton-shower
associated to a given event in the large-$N_c$ limit is the
following:
\begin{enumerate}
\item Calculate the $(n-1)!$ dual amplitudes corresponding to all
possible planar color configurations. This can be done
with {\tt ALPHA} using $N_c$ large enough \cite{Caravaglios:1999yr}.
\item Extract the {\em most likely} color configuration for this
event on a statistical basis, according to the relative contribution
of the single configurations to the total event
weight~\footnote{Defining $w_i=\vert A_i \vert^2$ for each color flow
$i$, and $W_i=\sum_{k=1,\dots,i} \, w_k/\sum_{k=1,\dots,n} \, w_k$,
the $j$-th color structure will be selected if $W_{j-1} \le \xi <
W_j$, for a random number $\xi$ uniformly distributed over the
interval $[0,1]$.}. Since each dual amplitude is gauge invariant, the
choice of color-configurations is also a gauge-invariant operation.
\item Develop the PS out of each initial and final-state
parton, starting from the selected color configuration. This step can
be carried out by feeding the generated event to a Monte-Carlo program
such as {\tt HERWIG}, which is precisely designed to {\em turn partons
into jets} starting from an assigned color-ordered configuration.
\end{enumerate}                                                        
Notice that, if the dual amplitudes are evaluated for a specific
helicity configuration, {\tt HERWIG} will also include
spin-correlation effects in the evolution of the parton
shower~\cite{Collins:1988cp,Knowles:1988vs,Marchesini:1988cf,Marchesini:1992ch,Marchesini:1996vc}.

As a result, use of the dual-amplitude representation of a multi-gluon
amplitude allows to accurately describe not only the large-angle
correlations in multi-jet final states, but also the full shower
evolution of the initial- and final-state partons with the same
accuracy available in {\tt HERWIG} for the description of 2-jet
final states.

In alternative to the above prescription, one can use {\tt ALPHA} to
calculate the matrix elements for external states with assigned
colors. Since these states are all orthogonal, such an approach is
particularly efficient if one wants to use a Monte Carlo method to
 sum over all possible color states.  The program will then
extract through a standard unweighting (at the leading order in
$1/N_c^2$) a specific color flow from all possible color flows
contributing to a given orthogonal color state.  This color flow is
then suitable as an initial condition for the shower evolution.
The advantage of this approach is twofold: first
the number of dual
amplitudes contributing to the amplitude for external states with assigned
colors is substantially smaller than the total one and
second  dual
amplitudes
are required only for {\em accepted events}
which, in general, are a small fraction of the generated ones.
Further details can be found in~\cite{Caravaglios:1999yr}. 

\subsection{Results}
As an example of our technique, we present here results for the
following two parton-level processes:
\begin{eqnarray}                     
g~g      &\to& 8~g  \nonumber \\
q~\bar q &\to& 8~g\,.
\end{eqnarray}
For comparisons, we also computed the above reactions in the 
simple approximation first suggested by Kunszt and Stirling~\cite{sphel}. 
This approximation (hereafter referred to as {\tt SPHEL}) consists in assuming
that the average value of maximally helicity
violating amplitudes \cite{Parke86}
is equal to             
the average value of all other non-zero amplitudes.

The kinematic configuration and the cut values used in our numerical examples
are as follows:
\ban
& \sqrt{\hat s} = 1500\,{\rm GeV}\,,  & \quad 
 p_{T_i} > 60\,{\rm GeV}\,, \quad |\eta_i| < 2\,
, 
\\
 & \Delta R_{ij} > 0.7\,. &
\ean                                          
These values, and the choice of a fixed strong coupling $\alpha_s= 0.12$,
only serve for illustrative purposes.

In Fig.~1, we show
the distribution of the minimum gluon transverse momentum for both
processes. 
In Fig~2 we plot the distributions for the
maximum gluon transverse momentum.

\begin{figure}
\centerline{\epsfig{figure= 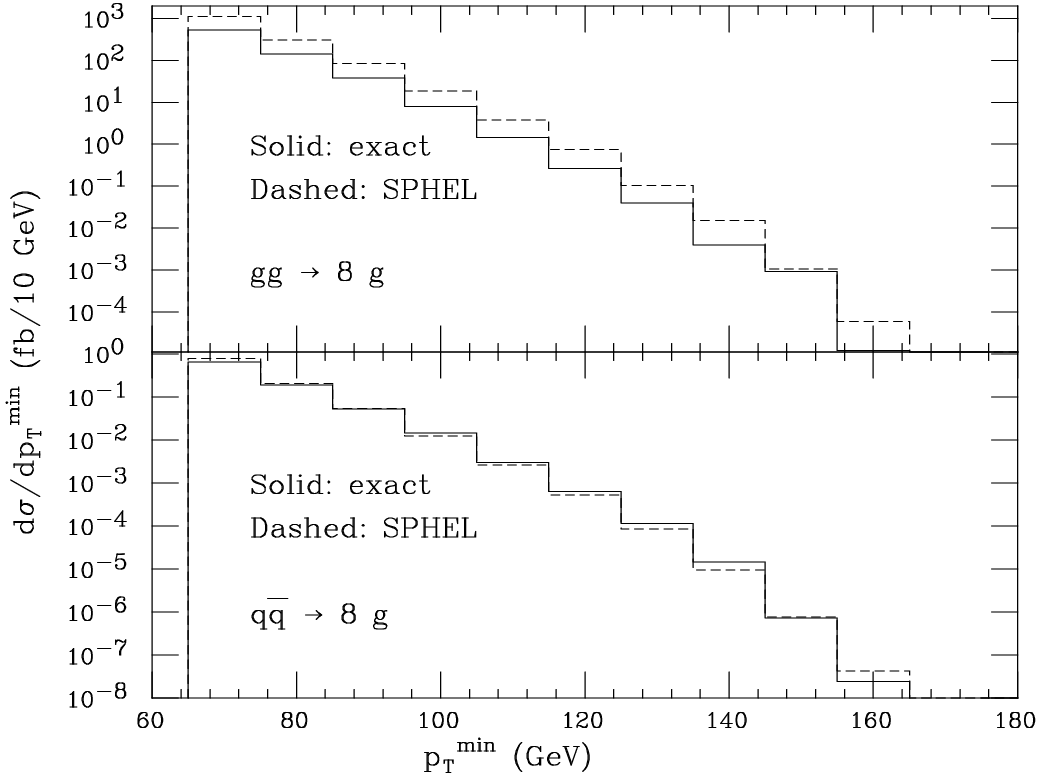, height= 7cm,width=7cm}}
\ccaption{}{Differential distributions for the minimum gluon 
transverse momentum. Exact result vs. {\tt SPHEL}.}
\end{figure}

\begin{figure}
\centerline{\epsfig{figure= 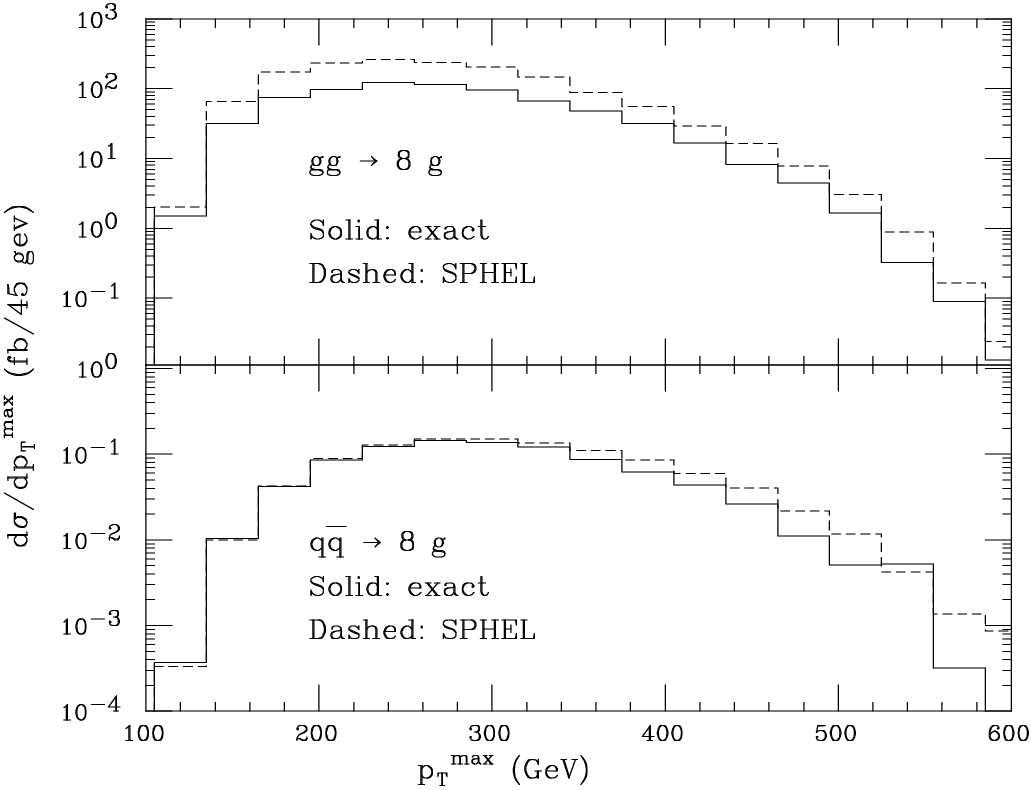, height= 7cm,width=7cm}}
\ccaption{}{Differential distributions for the maximun gluon 
transverse momentum. Exact result vs. {\tt SPHEL}.}
\end{figure}

\section{Conclusions}

We have reviewed
the {\tt ALPHA} algorithm to evaluate the exact, tree-level
matrix elements in the context of  multi-parton processes in QCD.
The algorithm is suitable (easy to implement) 
for the authomatic calculation of the scattering amplitudes
and it allows the inclusion of mass effects in a  straightforward manner. 
 The CPU cost  of the algorithm
grows like $K^n$ ($K$ constant, 
$n$ number of external particles)
as opposed to the factorial growth of the number of
Feynman graphs.

This technique
has been tested for processes such as $gg \to n$
gluons and $q\bar q \to n$ gluons, with $n$ up to 9. We discussed how the
summation over colour configurations  allows the construction of
parton-level event generators suitable to interfacing with a parton-shower
evolution including the effects of colour-coherence. This will eventually lead
to a fully exclusive, hadron-level description of multi-jet final states,
accurately incorporating the dynamics of large jet-jet separation angles.

\end{document}